\documentclass{ptapap}
\usepackage{amsmath}
\usepackage{amssymb}

\author{Fatemeh Hossein Nouri}[CPT]
\author{Agnieszka Janiuk}[CPT]

\affil[CPT]{Center for Theoretical Physics, Polish Academy of Sciences, Al. Lotnikow 32/46, 02-668 Warsaw, Poland}

\title{ACCRETION OF THE MAGNETIZED NEUTRINO-COOLED TORUS}

\begin{document}

\maketitle

\begin{abstract}

Neutrino-cooled accretion flow around a black hole, produced by a compact binary merger, is a promising scenario for jet formation and magnetic-driven winds to explain short duration gamma ray bursts (GRBs) central engine and kilonovae based on GW170817 gravitational wave observation.  Magnetorotational instability (MRI) turbulence and Blandford-Znajek (BZ) mechanism are expected to play key roles in the thermal equilibrium of the disk (balancing neutrino cooling) and in driving accretion and creating jets.  Using the open-source GRMHD HARM-COOL code, we study the magnetically-driven evolution of an accretion disk with realistic equation of state in the fixed curved space-time background. We identify the effects of the neutrino cooling and the magnetic field, paying particular attention to the dynamical, thermal and composition evolution of the disk and outflows.

\end{abstract}

\section{Introduction}

Given the hot, dense accretion flow in a post-merger black hole-accretion disk remnants, energy can be channeled into ultra-relativistic outflow needed to explain GRB properties,~\cite{Nakar2007,Berger:2013jza}. In such systems, the accretion gas cools by neutrino emission continuously,~\cite{Popham99,DiMatteo2002,Janiuk-2004}. On the other hand, magnetic fields can also extract energy from the disk and black hole spin, by generating the magnetic driven winds and Poynting flux-dominated jets,~\cite{BZ77}. Moreover, it can support the heating process of the plasma through the viscous effects of the MRI,~\cite{BalbusHaw1991}. 
The dynamical ejecta outflows during the merger, and the post-merger outflows are considered as possible sources explaining the observation of kilonova and other electromagnetic (EM) counterparts following the gravitational wave detections,~\cite{metzger:11,Dietrich-2017}. In this series of studies, we utilize the numerical simulations to investigate the evolution and different properties of the disk and the outflows in the presence of magnetic fields and neutrino cooling. 

\section{Numerical Methods and Initial Setups}

We evolve two analytical Fishbone-Moncrief disk models,~\cite{FM1976}, one with magnetic field and the other without magnetic field using the GRMHD HARM-COOL code,~\cite{Sap2019}. The metric is Kerr-Schild and frozen in time, and the hydro equations are evolved in conservative formalism using HLL shock capturing scheme. In this setup the black hole mass is set to $3M_{\odot}$, with spin $a=0.98$, and the disk size is set to have the total mass of $M_{disk}=0.1 M_{\odot}$. For the magnetized case, the initial magnetic field is confined within the torus by poloidal loops with the ratio of the maximum gas pressure to the maximum magnetic pressure $\beta = P_{gas,max}/P_{B,max}=50$. 

The equation of state is a realistic nuclear matter composed of free protons, neutrons, electron–positron pairs, and helium nuclei,~\cite{Reddy-1998}, considering weak interactions to determine neutrino opacities in hot dense matter. For neutrino treatment we use the formalism from~\cite{Janiuk-2013,Janiuk-2019}, where the neutrino cooling rate is given by

\begin{equation}
    Q_{\nu}^- = \frac{\frac{7}{8}\sigma T^4}{\frac{3}{4}} \sum_{i=e,\mu} \frac{1}{\frac{\tau_{a,\nu_i}+\tau_s}{2} + \frac{1}{\sqrt{3}} + \frac{1}{3 \tau_{a,\nu_i}}} \times \frac{1}{H}~[erg~s^{-1}~cm^{-3}],
\end{equation}

while the neutrino optical depth for different species are approximated as, ~\cite{DiMatteo2002}:

\begin{equation}
    \tau_{a,\nu_i} = \frac{H}{4\frac{7}{8}\sigma T^4} q_{a,\nu_i},
\end{equation}

where $q_{a,\nu_i}$ is the absorption rate for different neutrino species. 

These simulations were performed on the Prometheus cluster with 24 processors within a 2D grid with a resolution of 264*256 points on $r$ and $\theta$ in spherical coordinates. These models were evolved for $t=90$ms. 

\section{Results}

\begin{figure}
\centering
\begin{minipage}{0.48\textwidth}
\includegraphics[width=\textwidth]{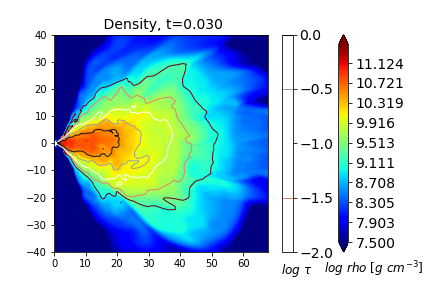}
\caption{The 2D profile of density and the contours of the optical depth of the magnetized disk with neutrino cooling at $t=0.03$s snapshot.}
\label{fig:rho-beta50}
\end{minipage}
\quad
\begin{minipage}{0.48\textwidth}
\includegraphics[width=\textwidth]{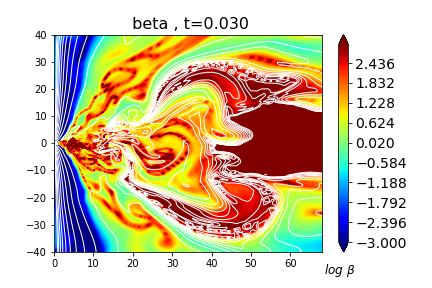}
\caption{The 2D profile of the ratio of gas pressure to magnetic pressure ($\beta$) and magnetic field contour lines of the magnetized disk with neutrino cooling at $t=0.03$s snapshot.}
\label{fig:beta-beta50}
\end{minipage}
\end{figure}

\begin{figure}
\centering
\begin{minipage}{0.48\textwidth}
\includegraphics[width=\textwidth]{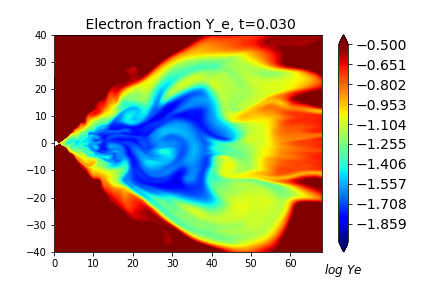}
\caption{The 2D profile of the electron fraction ($Y_e$) at $t=0.03$s snapshot.}
\label{fig:ye-beta50}
\end{minipage}
\quad
\begin{minipage}{0.48\textwidth}
\includegraphics[width=\textwidth]{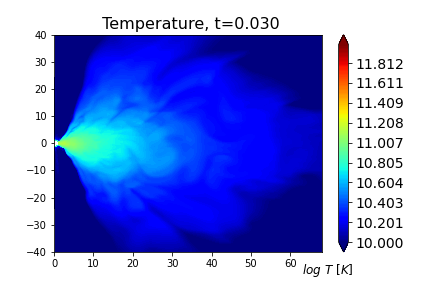}
\caption{The 2D profile of the temperature at $t=0.03$s snapshot.}
\label{fig:temp-beta50}
\end{minipage}
\end{figure}

The results of the magnetized simulation are presented in Figs.~\ref{fig:rho-beta50}-\ref{fig:temp-beta50} at $t=0.03$s time snapshot. The 2D profiles show that disk becomes turbulent and expands from the outer boundary. There is a significant accretion happening as a result of MRI, and the disk is quite transparent for neutrino emissions. The funnel shape jet structure with high magnetic field is formed at the early time of the evolution. Measuring the volume density-averaged electron fraction shows that the disk becomes less neutron-rich and less degenerate at the end of the simulation. 

\begin{figure}
\centering
\begin{minipage}{0.48\textwidth}
\includegraphics[width=\textwidth]{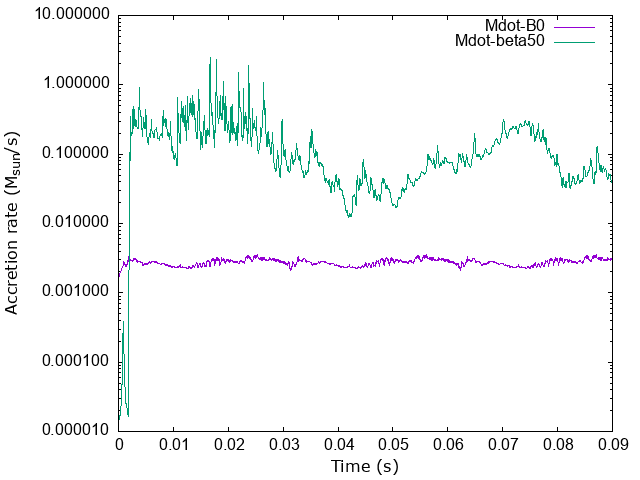}
\caption{The accretion rate comparison between magnetized and non-magnetized cases.}
\label{fig:mdot}
\end{minipage}
\quad
\begin{minipage}{0.48\textwidth}
\includegraphics[width=\textwidth]{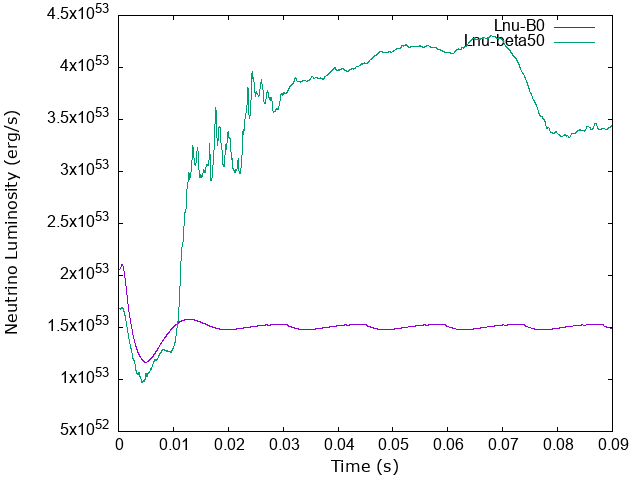}
\caption{The neutrino luminosity comparison between magnetized and non-magnetized cases.}
\label{fig:lnu}
\end{minipage}
\end{figure}

The comparison between the magnetized and non-magnetized cases are given in Figs.~\ref{fig:mdot} and~\ref{fig:lnu}. As one expects, MRI provides the required mechanism for the angular momentum transport and matter accreting into the black hole. On the other hand, we observe that the neutrino luminosity increases by factor of four for the magnetized case during the evolution. This can be explained as a thermal competition between magnetic-driven viscous heating effect and the neutrino and advection cooling effects, which shifts the equilibrium condition for the weak interactions and generates more neutrinos.  

\begin{figure}
\centering
\begin{minipage}{0.48\textwidth}
\includegraphics[width=\textwidth]{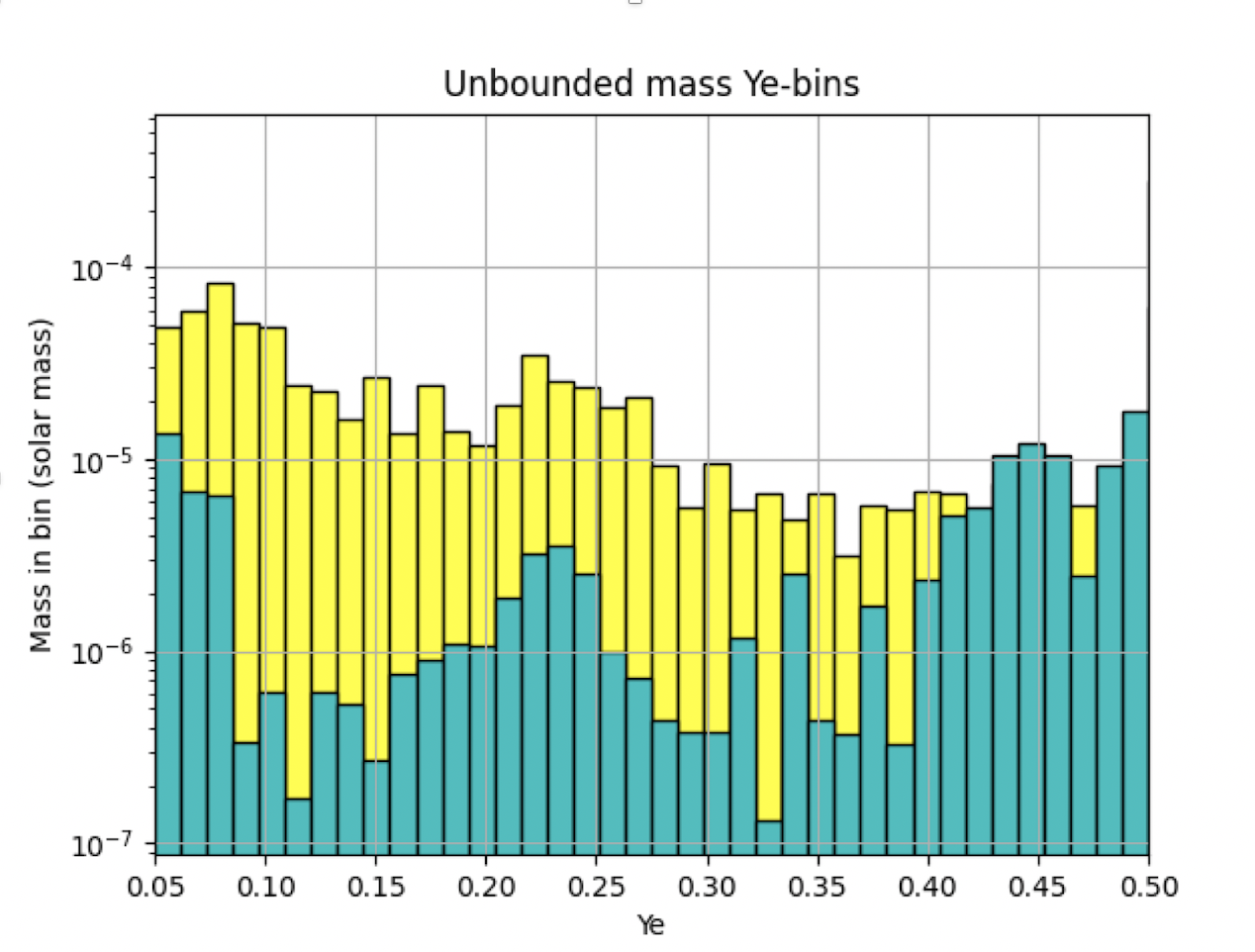}
\caption{Histograms of unbounded mass as a function of electron fraction. The yellow is the average taken over several time snapshots from the first half of the evolution time, while the cyan is the average from the second half.}
\label{fig:unbound-ye}
\end{minipage}
\quad
\begin{minipage}{0.48\textwidth}
\includegraphics[width=\textwidth]{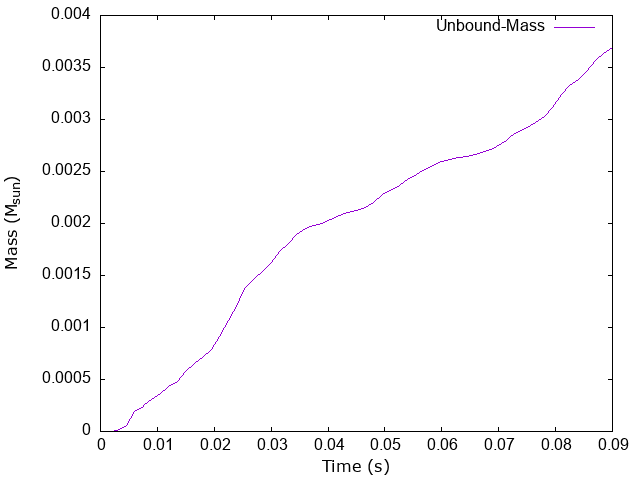}
\caption{The unbound mass generated during the evolution. The rate of the unbound mass decreases by time, because it is not possible to maintain the magnetic field in 2D for a long evolution in order to support the angular momentum transport via MRI.}
\label{fig:unbound-mass}
\end{minipage}
\end{figure}

\begin{figure}
\centering
\begin{minipage}{0.48\textwidth}
\includegraphics[width=\textwidth]{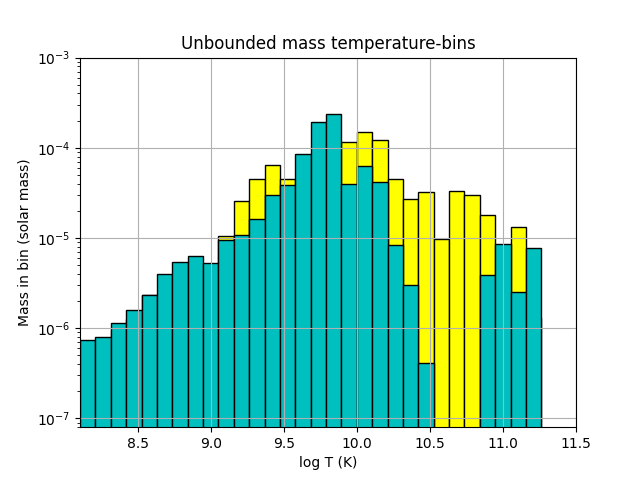}
\caption{Histograms of unbounded mass as a function of temperature. The averages are taken same as Fig.~\ref{fig:unbound-ye}.}
\label{fig:unbound-temp}
\end{minipage}
\quad
\begin{minipage}{0.48\textwidth}
\includegraphics[width=\textwidth]{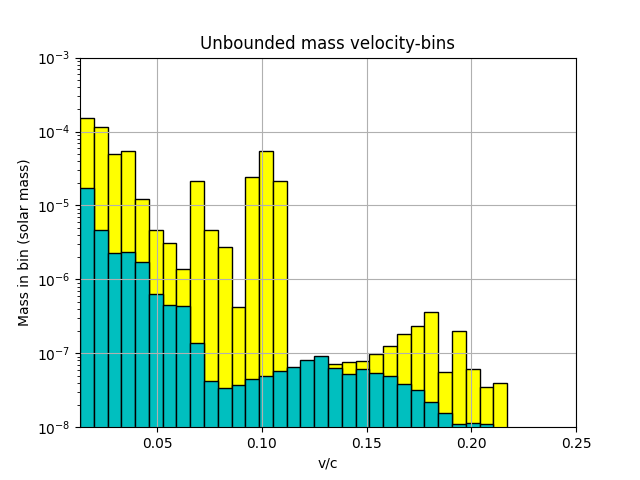}
\caption{Histograms of unbounded mass as a function of velocity. The averages are taken same as Fig.~\ref{fig:unbound-ye}.}
\label{fig:unbound-v}
\end{minipage}
\end{figure}

For the magnetized case, we mark the fluid with positive energy $-hu_t > 1$ as unbound matter (here $h$ is the enthalpy and $u_t$ is the $t$ component of the 4-velocity). Taking summation over density of the unbound matter for the entire grid, we plot the histograms to measure the total mass over the electron fraction $Y_e$, temperature and velocity bins. These analysis are done at two time intervals, by taking the average over several snapshots from the first half of the evolution, and the same for the second half. These histograms and the mass of the unbound matter verses time are illustrated in Figs.~\ref{fig:unbound-ye}-\ref{fig:unbound-v}.
Overall, the outflows become colder with lower kinetic energy at the end of the simulation. This observation can be explained by the fact that the magnetic field dissipates over time in a two dimensional evolution as a result of the anti-dynamo effect. Hence, the generation of the magnetic-driven winds slows down by further evolution.

\section{Summary and Conclusions}

We evolved two analytical disk-black hole models with and without magnetic field. We observe the creation of the jets and turbulent plasma as a result of BZ mechanism and MRI. The neutrino cooling has a significant effect on the energy budget of the disk and its thermal and composition evolution. Measuring the unbounded mass, and its features such as temperature, velocity and composition, shows that our model does not produce enough energetic outflows (yet) to compete with the dynamical ejecta to explain the source of the EM counterparts (see~\cite{Dietrich-2017} for the outflow estimates for black hole-neutron star and binary neutron star merger simulations). We plan to perform more simulations with different setup such as different spin and different mass ratios for disks and black holes to investigate their effects on the outflows. 3D simulations and more accurate approximation for neutrino treatment such as neutrino leakage scheme are required for future studies to estimate the outflows' properties more precisely. 

\acknowledgements{
This research was supported in part by grant No. 2019/35/B/ST9/04000 from the Polish National Science Center.
}

\bibliographystyle{ptapap}
\bibliography{hossein-nouri}

\begin{thebibliography}{14}
\providecommand{\natexlab}[1]{#1}
\providecommand{\url}[1]{\texttt{#1}}
\providecommand{\urlprefix}{URL }
\providecommand{\eprint}[2][]{\url{#2}}

\bibitem[{{Balbus} \& {Hawley}(1991)}]{BalbusHaw1991}
{Balbus}, S.~A., {Hawley}, J.~F., \emph{ApJ} \textbf{376}, 214 (1991)

\bibitem[{Berger(2014)}]{Berger:2013jza}
Berger, E., \emph{Ann. Rev. Astron. Astrophys.} \textbf{52}, 43 (2014)

\bibitem[{{Blandford} \& {Znajek}(1977)}]{BZ77}
{Blandford}, R.~D., {Znajek}, R.~L. \textbf{179}, 433 (1977)

\bibitem[{{Di Matteo} et~al.(2002){Di Matteo}, {Perna}, \&
  {Narayan}}]{DiMatteo2002}
{Di Matteo}, T., {Perna}, R., {Narayan}, R., \emph{\apj} \textbf{579}, 2, 706
  (2002)

\bibitem[{{Dietrich} \& {Ujevic}(2017)}]{Dietrich-2017}
{Dietrich}, T., {Ujevic}, M., \emph{Classical and Quantum Gravity} \textbf{34},
  10, 105014 (2017)

\bibitem[{{Fishbone} \& {Moncrief}(1976)}]{FM1976}
{Fishbone}, L.~G., {Moncrief}, V., \emph{\apj} \textbf{207}, 962 (1976)

\bibitem[{{Janiuk}(2019)}]{Janiuk-2019}
{Janiuk}, A., \emph{\apj} \textbf{882}, 2, 163 (2019)

\bibitem[{{Janiuk} et~al.(2013){Janiuk}, {Mioduszewski}, \&
  {Moscibrodzka}}]{Janiuk-2013}
{Janiuk}, A., {Mioduszewski}, P., {Moscibrodzka}, M., \emph{\apj} \textbf{776},
  2, 105 (2013)

\bibitem[{{Janiuk} et~al.(2004){Janiuk}, {Perna}, {Di Matteo}, \&
  {Czerny}}]{Janiuk-2004}
{Janiuk}, A., {Perna}, R., {Di Matteo}, T., {Czerny}, B., \emph{Mon. Not. Roy.
  Astron. Soc.} \textbf{355}, 950 (2004)

\bibitem[{{Metzger} \& {Berger}(2012)}]{metzger:11}
{Metzger}, B.~D., {Berger}, E. \textbf{746}, 48 (2012)

\bibitem[{{Nakar}(2007)}]{Nakar2007}
{Nakar}, E., \emph{Phys. Rep.} \textbf{442}, 166 (2007)

\bibitem[{Popham et~al.(1999)Popham, Woosley, \& Fryer}]{Popham99}
Popham, R., Woosley, S.~E., Fryer, C., \emph{Astrophys. J.} \textbf{518}, 356
  (1999)

\bibitem[{Reddy et~al.(1998)Reddy, Prakash, \& Lattimer}]{Reddy-1998}
Reddy, S., Prakash, M., Lattimer, J.~M., \emph{Phys. Rev. D} \textbf{58},
  013009 (1998),
  \urlprefix\url{https://link.aps.org/doi/10.1103/PhysRevD.58.013009}

\bibitem[{{Sapountzis} \& {Janiuk}(2019)}]{Sap2019}
{Sapountzis}, K., {Janiuk}, A., \emph{\apj} \textbf{873}, 1, 12 (2019)

\end{thebibliography}

\end{document}